\documentclass[twocolumn,aps,prl,reprint]{revtex4-1}
\usepackage{amsmath}
\usepackage{amssymb}
 \usepackage[toc,page]{appendix}
\usepackage[usenames,dvipsnames]{xcolor}
\usepackage[normalem]{ulem}

\usepackage{tikz}
\usetikzlibrary{decorations.markings}
\usepackage{braket}
\usepackage{amsfonts}
\newcommand{\Tr}{\text{Tr}}
\usepackage{verbatim}
\usepackage{subcaption}
\usepackage{graphicx}
\usepackage{tikz}
\usepackage{lipsum}
\usepackage{float}

\begin{document}

\title{Entanglement Amplification from Rotating Black Holes}
\author{Matthew P. G. Robbins${}^{1,2,3}$, Laura J. Henderson${}^{1,4}$, and Robert B. Mann${}^{1,2,3,4}$}
\affiliation{Department of Physics and Astronomy, University of Waterloo, Waterloo ON, Canada, N2L 3G}
\affiliation{Perimeter Institute for Theoretical Physics, 31 Caroline Street North, Waterloo ON, Canada, N2L 2Y5}
\affiliation{Waterloo Centre for Astrophysics, University of Waterloo, Waterloo ON, Canada, N2L 3G1}
\affiliation{Institute for Quantum Computing, University of Waterloo, Waterloo ON, Canada, N2L 3G}

\begin{abstract}
The quantum vacuum has long been known to be characterized by field   correlations between spacetime points.  These correlations can be swapped with a pair of  
particle detectors, modelled as simple  two-level quantum systems (Unruh-DeWitt detectors) via a process known as entanglement harvesting.   We study this phenomenon in the presence of a rotating BTZ black hole, and find that rotation can significantly amplify the harvested vacuum entanglement. Concurrence between co-rotating detectors is amplified by as much as an order of magnitude at intermediate distances from the black hole relative to that at large distances.  The effect is most pronounced for  near-extremal small mass black holes, and allows for harvesting at large spacelike detector separations.  We also find that the entanglement shadow -- a region near the black hole from which entanglement cannot be extracted -- is diminished in size as  the black hole's angular momentum increases. 
\end{abstract}

\maketitle
\noindent{\it Introduction}\label{sec: into} $\quad$
The natural entanglement of the vacuum in quantum field theory  plays a crucial role in multiple areas of physics, including black hole entropy \cite{Solodukhin2011,Brustein2005}, the AdS/CFT correspondence \cite{Ryu2006}, quantum information \cite{Peres2004,Lamata1997},  
and metrology \cite{Ralph2009}. Quantum vacuum correlations have long been known to be between both timelike and spacelike separated regions~\cite{summers1985bell,summers_bells_1987}, and are at the heart of
the black hole information paradox~\cite{Preskill:1992tc,Mathur:2009hf} and its proposed solutions~\cite{Almheiri:2012rt,Braunstein:2009my,Mann:2015luq}, as well as  playing a key role in quantum energy teleportation~\cite{doi:10.1143/JPSJ.78.034001,Hotta:2011xj}.  It was  later realized   \cite{Valentini1991} that    vacuum entanglement can be swapped with 
a physical system: two initially uncorrelated atoms (either spacelike or timelike separated) interacting with the electromagnetic vacuum for a finite time can exhibit nonlocal correlations. 
The process is best explicated in an idealized system of two qubits (modelled as Unruh-DeWitt (UDW) detectors  \cite{Unruh1976,deWitt}) interacting
with a scalar field \cite{Reznik2003,Reznik2005}. A protocol known as entanglement harvesting  \cite{Salton2015} was then developed 
in which the entanglement in the scalar quantum vacuum is transferred to the UDW detectors. This phenomenon has proven to be very useful in
characterizing properties of the quantum vacuum that are not accessible to a single detector, including the thermal character of de Sitter spacetime
\cite{PhysRevD.79.044027,Huang2017}, probing spacetime topology \cite{Smith2016},  and finding new structures such as separability islands in anti-de Sitter 
spacetime \cite{Ng:2018drz,Henderson2019}.

Surprisingly little is known about the extraction of vacuum entanglement  for spacetimes with black holes.  
 The first such study
 \cite{Henderson2018} indicated that black holes have entanglement shadows: a region outside the black hole within which it is not possible to
 harvest entanglement.  Although the study was carried out for the $(2+1)$-dimensional Banados-Teitelboim-Zanelli (BTZ) black hole \cite{Banados1992}, the phenomenon is
 expected to be universal \cite{Henderson:2019uqo}.  A recent study of entanglement harvesting   in Schwartzschild/Vaidya spacetimes is commensurate with this expectation \cite{Tjoa2020}.
 
 We present  here the results of the first study of entanglement harvesting for rotating black hole spacetimes.  Concentrating specifically on a rotating BTZ black hole,  we find the remarkable result  that rotation markedly amplifies the harvested entanglement:   as much as a 10-fold increase in the concurrence between 2 UDW detectors at proper distances as far as 100 horizon radii from the black hole is possible. The effect is most dramatic for near-extremal small-mass black holes, and allows for harvesting at large spacelike detector separations that would otherwise yield zero concurrence.  We also find that the entanglement shadow diminishes with increasing angular momentum and that the concurrence  does not monotonically increase with increasing energy gap, in contrast to the static case.

{\it  Entanglement Harvesting Protocol}\; 
\label{sec: fundamentals}
The  light-matter interaction, assuming no angular momentum exchange, is well described as a local interaction between  a scalar quantum field $\phi(x)$
and a UDW detector (whose respective ground $\ket{0}_D$ and excited $\ket{1}_D$ states are separated by an energy gap $\Omega_D$) moving along a spacetime trajectory $x_D(\tau)$ \cite{Funai:2018wqq}.  The interaction Hamiltonian is
\begin{align}
H_D=\lambda\chi_D(\tau)\left(e^{i\Omega\tau}\sigma^+ + e^{-i\Omega\tau}\sigma^-\right)\otimes\phi[x_D(\tau)]
\end{align}
where $\chi_D(\tau) \leq 1 $ is a switching function controlling the duration of the interaction,
$\lambda\ll1$ is the field/detector coupling constant, and $\sigma^+= \ket{1}_D\bra{0}_D$, $\sigma^-= \ket{0}_D\bra{1}_D$ are ladder operators that raise and lower the energy levels of the UDW detectors.  

Suppose we have two detectors $A$ and $B$ with trajectories $x_A(\tau_A)$ and $x_B(\tau_B)$. If the initial state of the detector-field system is $\ket{\Psi_i}=\ket{0}_A\ket{0}_B\ket{0}$, then after a time $t$, $\ket{\Psi_f}=U(t,0)\ket{\psi_i}$, where $U(t,0)=\mathcal{T}e^{-i\int dt\left[\frac{d\tau_A}{dt}H_A(\tau_A)+\frac{d\tau_B}{dt}H_B(\tau_B)\right]}$, with $\mathcal{T}$  the time-ordering operator. Letting $\rho_{AB}=\Tr_\phi\ket{\Psi_f}\bra{\Psi_f}$ be the reduced density operator describing the detectors after tracing over the scalar field degrees of freedom, we have \cite{Smith2016,Smith:2017vle}
\begin{align}
\rho_{AB}=\begin{pmatrix}
1-P_A-P_B& 0 & 0   & X   \\
0  & P_B  &  L_{AB}& 0  \\
 0 & L_{AB}^* &  P_A &0  \\
  X^* & 0& 0  &   0
\end{pmatrix}
+\mathcal{O}(\lambda^4)
\end{align}
in the basis $\{\ket{0}_A\ket{0}_B,\ket{0}_A\ket{1}_B,\ket{1}_A\ket{0}_B,\ket{1}_A\ket{1}_B\}$,  where $P_D$ is the probability for detector $D$ ($D=A,B$) to become excited, the non-local corrections are denoted by $X$, and $L_{AB}$ describes the non-entangling correlations. See the Supplementary Material for explicit calculations of $P_D$, $X$, and $L_{AB}$. These quantities are all dependent on the two-point correlation function, $W(x,x')=\braket{0|\phi(x)\phi(x')|0}$ (also called the Wightman function) of the vacuum. The entanglement present in the system can be quantified using the concurrence \cite{Smith2016,Smith:2017vle,Wooters1998}
\begin{align}
\mathcal{C}\left[\rho_{AB}\right]=2\max\left[0,|X|-\sqrt{P_AP_B}\right]+\mathcal{O}(\lambda^4)
\end{align}
which qualitatively is non-zero when non-local detector correlations are greater than the geometric mean of detector noise.

{\it Entanglement Harvesting around a Rotating BTZ Black Hole}\; 
\label{sec: BTZ}

We are interested in implementing the entanglement harvesting protocol near a rotating BTZ black hole, whose line element is  \cite{Banados1992}
\begin{equation}\label{btzmet}
ds^2=-(N^\perp)^2dt^2+f^{-2}dr^2+r^2(d\phi+N^{\phi}dt)^2
\end{equation}
where, $N^\perp=f=\sqrt{-M+\frac{r^2}{\ell^2}+\frac{J^2}{4r^2}}$ and $N^\phi =-\frac{J}{2r^2}$ with $M=\frac{r_+^2+r_-^2}{\ell^2}$  and $J=\frac{2r_+ r_-}{\ell}$ the respective mass and angular momentum of the black hole, whose respective inner and outer horizon radiii are $r_-$ and $r_+$; $\ell$ is the AdS length. Note that   $|J|\leq M\ell$, with equality yielding extremality ($r_+=r_-$).

For a conformally coupled scalar field (in the Hartle-Hawking vacuum) the Wightman function is known analytically \cite{Lifschytz1994,Carlip1998}, and
can be written as the image sum
\begin{align}
W_{BTZ}(x,x')=\sum_{n=-\infty}^\infty \eta^nW_{AdS_3}(x,\Gamma^nx')
\end{align}
over the vacuum Wightman functions for AdS${}_3$, where $\Gamma x'$ takes $(t,r,\phi)\to(t,r,\phi+2\pi)$ and $\eta=\pm 1$ describes the untwisted/twisted nature of the scalar field. With this identification, we have \cite{Hodgkinson2012,Smith:2013zqa}
\begin{widetext}
\begin{align}
W_{BTZ}=\frac{1}{4\pi}\frac{1}{2\sqrt{\ell}}\sum_{n=-\infty}^{n=\infty}\eta^n\left(\frac{1}{\sqrt{\sigma_{\epsilon}(x,\Gamma^nx')}}-\frac{\zeta}{\sqrt{\sigma_{\epsilon}(x,\Gamma^nx')+2}}\right)
\label{eq: sum}
\end{align}
where
\begin{equation}
\begin{aligned}
\sigma_\epsilon(x,\Gamma^nx')^2=&-1+\sqrt{\alpha(r)\alpha(r')}\cosh\left[\frac{r_+}{\ell}(\Delta\phi-2\pi n)-\frac{r_-}{\ell^2}(t-t')\right]\\
&-\sqrt{(\alpha(r)-1)(\alpha(r')-1)}\cosh\left[\frac{r_+}{\ell^2}(t-t')-\frac{r_-}{\ell}(\Delta\phi-2\pi n)\right]
\label{eq: sigma 0}
\end{aligned}
\end{equation}
\end{widetext}
and 
\begin{align}
\alpha(r)&=\frac{r^2-r_-^2}{r_+^2-r_-^2}\qquad
\Delta\phi=\phi-\phi'\ .
\end{align}
The boundary conditions for the scalar field (Dirichlet, Neumann, transparent) are represented by $\zeta=1,-1,0$, respectively. We shall consider detectors with equal energy gaps $\Omega_A=\Omega_B=\Omega$ and switching functions $\chi(\tau_A)=e^{-\tau_A^2/2\sigma^2}$ and $\chi(\tau_B)=e^{-\tau_B^2/2\sigma^2}$ (see Supplementary Material). We also set the AdS length to be $\ell/\sigma=10$ and only consider untwisted scalar fields with $\eta=1$ with Dirichlet boundary conditions of $\zeta=1$. We obtain qualitatively similar results for $\zeta=-1$ and $0$.




We consider two detectors $A$ and $B$, respectively located at $R_A$ and $R_B$
 whose proper separation is
\begin{align}
d(R_A,R_B)=\ell\log
   \left(\frac{\sqrt{R_B^2-r_-^2}+\sqrt{R_B^2-r_+^2}}{\sqrt{R_A^2-r_-^2}+\sqrt{R_A^2-r_+
   ^2}}\right)
\end{align}
at fixed $(t,\phi)$. The proper time $\tau_D$ for each detector is related to these coordinates via
\begin{align}
\phi_D &=  \frac{r_{-}} {\ell r_+} t_D =  \frac{  r_-\tau_D }{\sqrt{R_D^2-r_+^2}\sqrt{r_+^2-r_-^2}}   
 \label{eq: CRM}
\end{align}
in the co-rotating frame \cite{Hodgkinson2012} for $D=A,B$. We keep 
 $d(R_A,R_B)/\sigma$ fixed whilst computing the concurrence for as a function of  the proper distance  $d(r_+,R_A)/\sigma$ between detector $A$ and the outer horizon at $r=r_+$.

\begin{figure*}[htb!]
        \begin{subfigure}{\textwidth}
\centering
\includegraphics[scale=0.9]{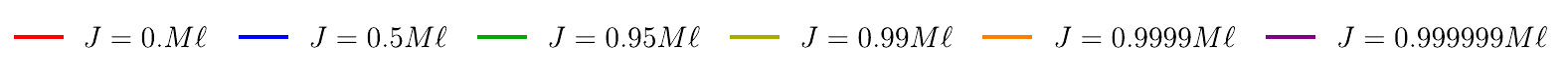}
\end{subfigure}
\begin{subfigure}{\textwidth}
\foreach \y in {1,2,3}
{
  \centering
    \includegraphics[width=0.32\linewidth]{M=0,001_plotCAll_Omega=\y_zeta=1.pdf}
}
\end{subfigure}
\caption{Concurrence of two UDW detectors separated by a distance $d(R_A,R_B)=1$ orbiting a black hole of mass $M=10^{-3}$ for various angular momenta. We have set $\zeta=1$ and $\ell/\sigma=10$.
}
\label{fig: ConcurrenceJ}
\end{figure*}

Our main result is illustrated in figure~\ref{fig: ConcurrenceJ}, which shows that rotation has a profound effect on the entanglement that can be harvested.
For small gap and small angular momenta, the concurrence is a monotonically increasing function of proper separation of detector $A$ from the horizon, until it asymptotes to its 
value in AdS spacetime, independent of both the angular momentum and mass, which we have verified numerically.
As the angular momentum of the black hole increases, there is relatively little change from the $J=0$ case.  However departures become apparent once $J/M\ell \geq 0.9$: we see  that the concurrence slowly grows and (from the insets) that the  entanglement shadow shrinks in size. The growth in the concurrence becomes quite rapid as the black hole approaches extremality, peaking at a value 8 times as large as the non-rotating case for $\Omega\sigma = 0.01$. The peak occurs quite far from the horizon, at about $d(r_+,R_A) = 25\sigma$, or about 100 horizon radii.  

As the gap $\Omega$ increases, these trends are exaggerated. The entanglement shadow shrinks further and the maximal concurrence near extremality grows larger, becoming  as much as 10 times greater than the $J=0$ case for $\Omega\sigma = 0.1$. 
  The growth in $\mathcal{C}$ becomes even more rapid and the shadow continues to decrease in size as $\Omega \to \sigma$, but 
the maximal concurrence  begins to diminish slightly, becoming 4 times larger than the $J=0$ case.  As  $d(r_+,R_A)$ become large, the concurrence decreases, asymptoting to its AdS value regardless of the value of $J/M\ell$. We find that for smaller values of  $J$ there is both enhancement and diminishment in the concurrence before asymptoting to the AdS value at large distances, whose detailed structure we will investigate in future work.
\begin{figure}[h!]
\includegraphics[width=\columnwidth]{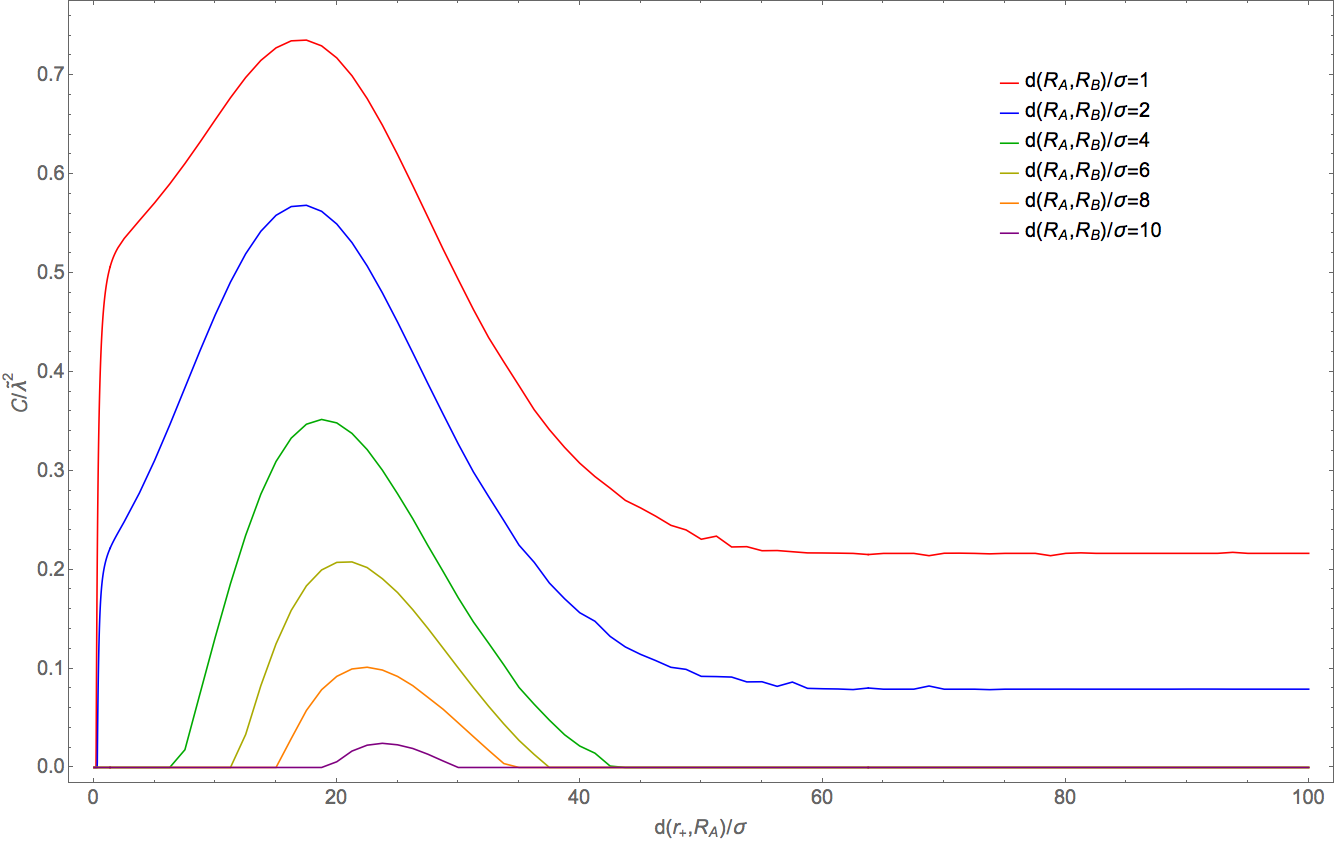}
\caption{Concurrence as a function of $d(r_+,R_A)/\sigma$ for various co-rotating detector separations $d(R_A,R_B)$, for $M=10^{-3}$,  $\zeta=1$, $\ell/\sigma=10$, $\Omega\sigma=1$, and $J/M\ell=0.9999$.}
\label{fig: proper distance}
\end{figure}

The effect diminishes as proper separation between the detectors increases, though still persists at intermediate distances (up to a maximum proper separation), as shown in Figure \ref{fig: proper distance} for $\Omega\sigma=1$. We find it is possible to extract entanglement at what are effectively spacelike
detector separations where the overlap between the switching functions is $\lesssim 10^{-8}$; indeed entanglement can be extracted for separations as large 
 $d(R_A,R_B)=10\sigma$, where the overlap is $\lesssim 10^{-22}$.   
This is quite remarkable -- small mass near-extremal black holes allow for entanglement extractions at large detector separations that would otherwise not be possible. 

\begin{figure*}[t]
    \begin{subfigure}{\textwidth}
\centering
\includegraphics[scale=0.85]{AllLegend.pdf}
\end{subfigure}
\centering
    \begin{subfigure}{0.3\textwidth}
  \centering
    \includegraphics[width=\textwidth]{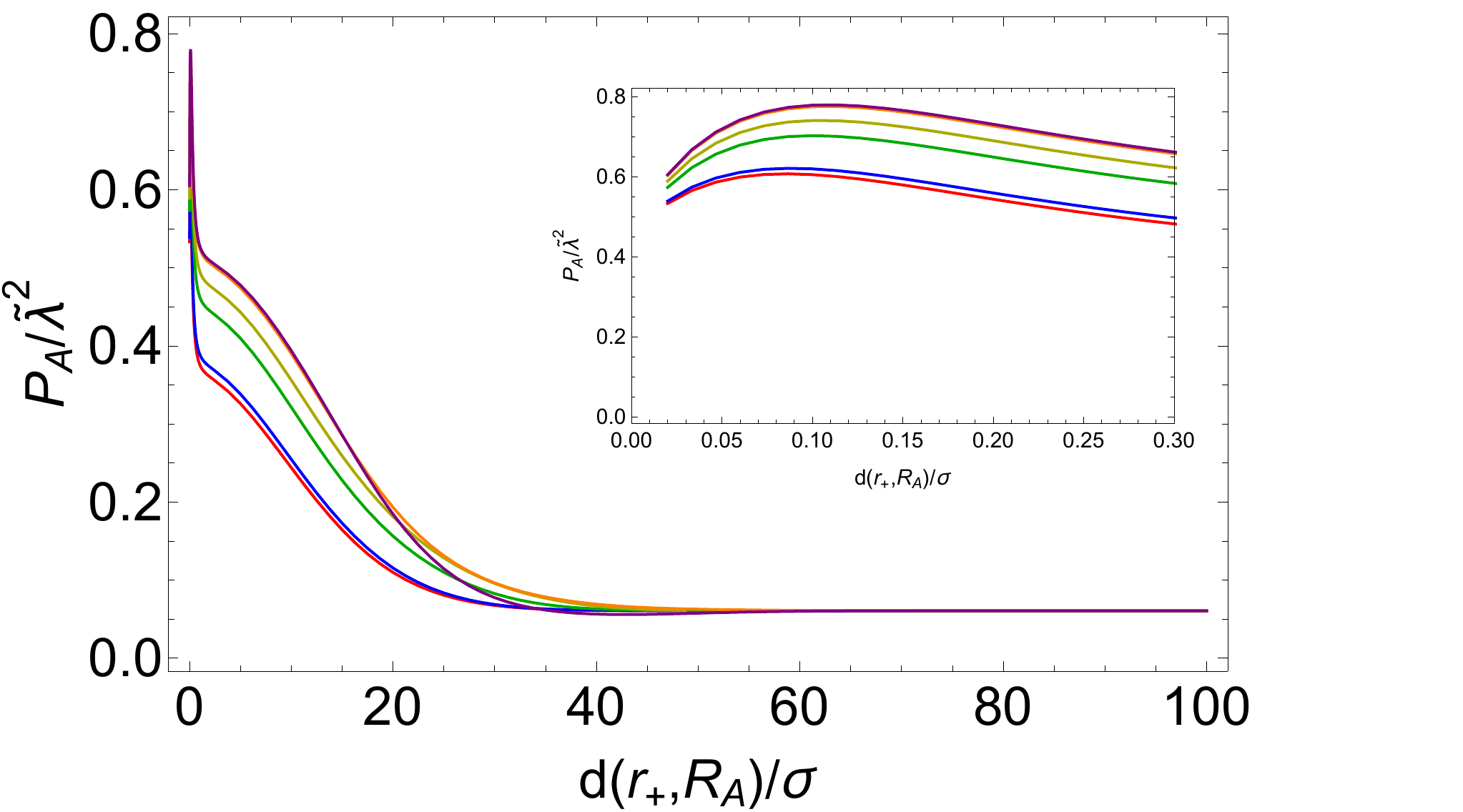}
    \caption{$P_A$}
    \end{subfigure}
\begin{subfigure}{0.3\textwidth}
  \centering
    \includegraphics[width=\textwidth]{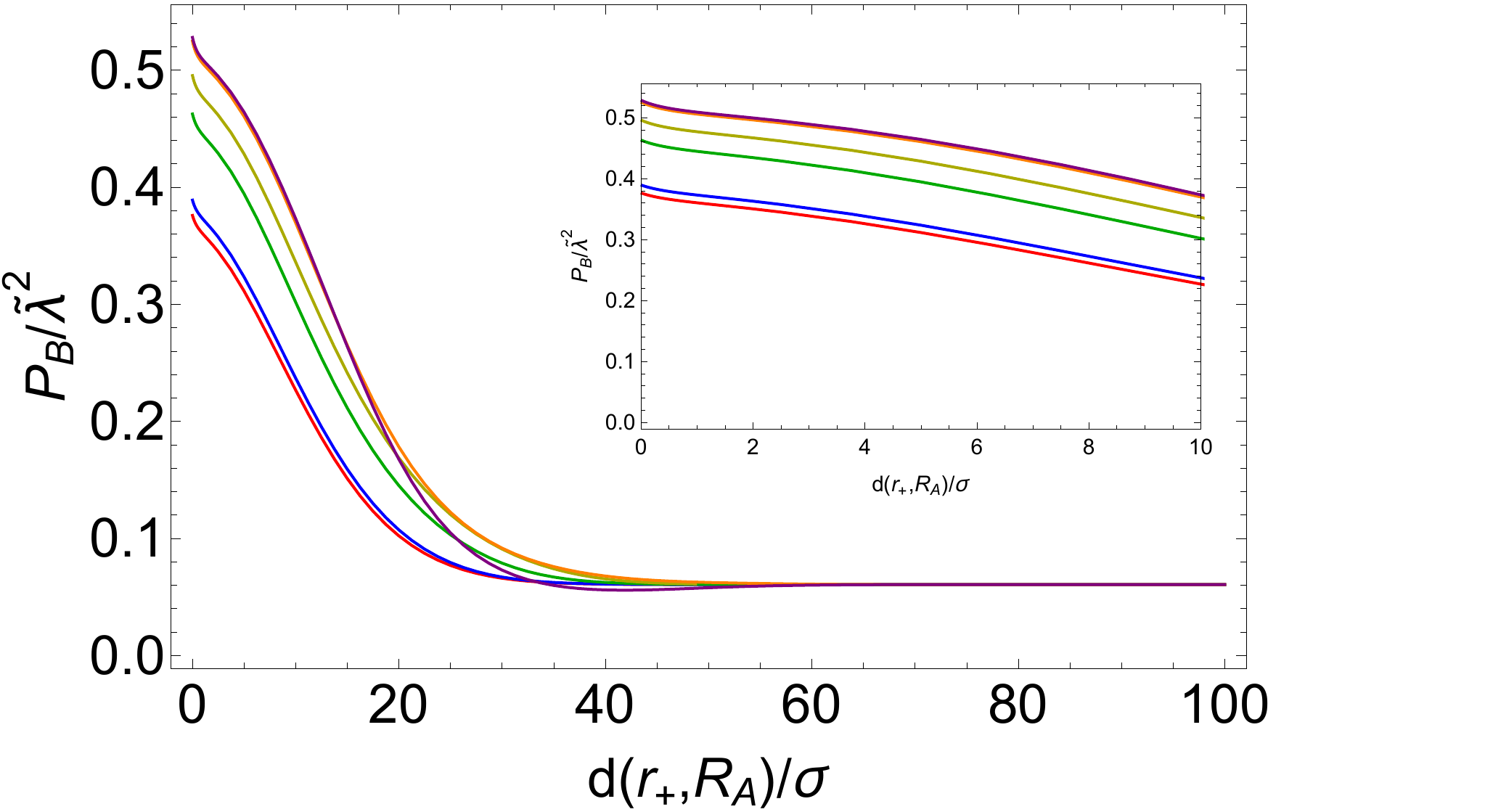}
    \caption{$P_B$}
    \end{subfigure}
    \begin{subfigure}{0.3\textwidth}
  \centering
    \includegraphics[width=\textwidth]{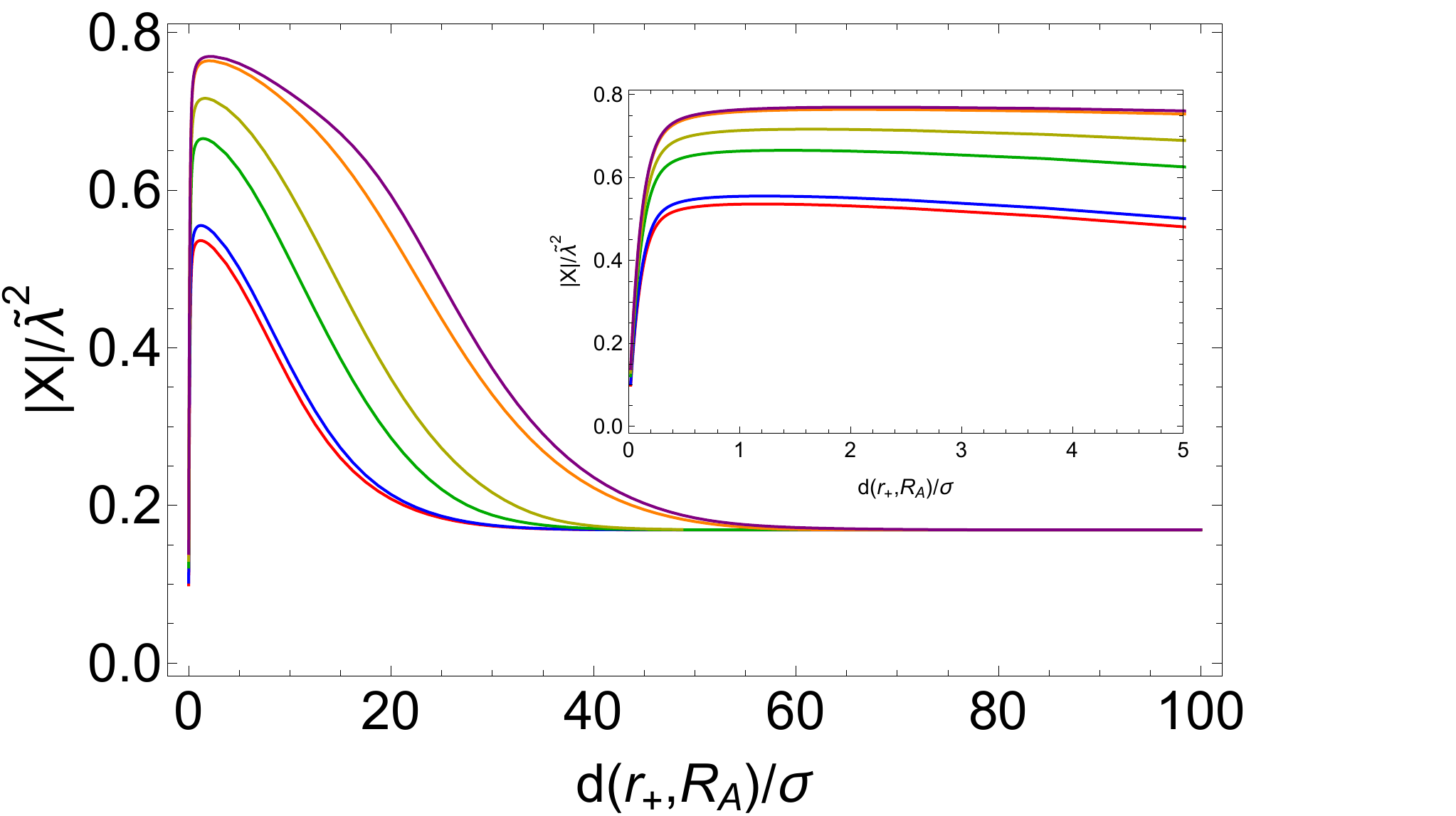}
    \caption{$|X|$}
    \end{subfigure}
           \begin{subfigure}{0.3\textwidth}
  \centering
    \includegraphics[width=\textwidth]{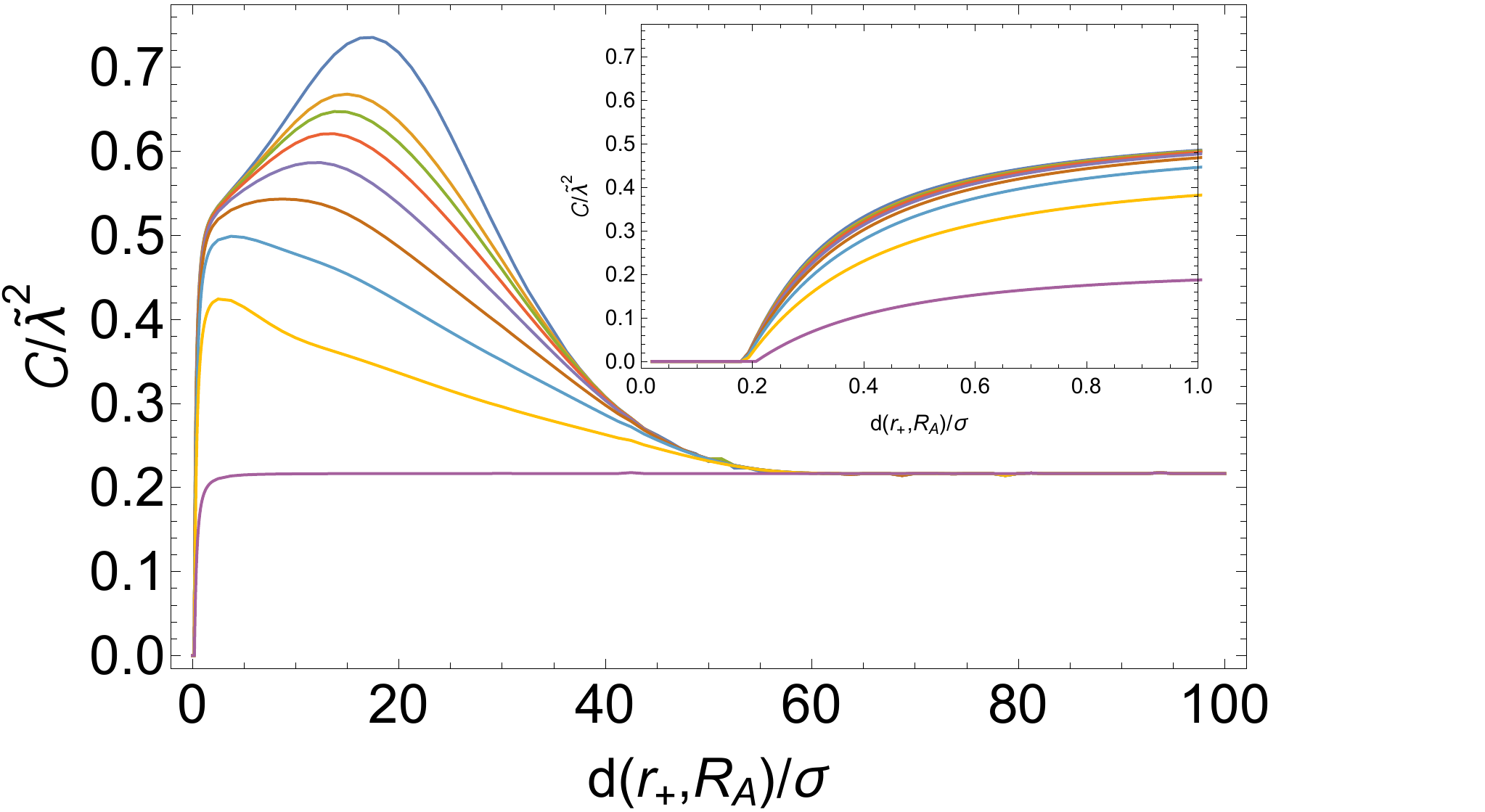}
    \caption{$\mathcal{C}$}
    \label{fig: partial sum}
    \end{subfigure}
        \hspace*{-0.8cm}
               \begin{subfigure}{0.01\textwidth}
               \vspace*{-0.8cm}
  \centering
    \includegraphics[scale=0.5]{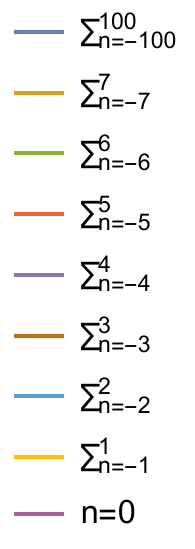}
    \end{subfigure}
        \caption{(a-c) Transition probabilities and non-locality for UDW detectors orbiting a rotating BTZ black hole. (d) Partial image sums of the concurrence for UDW detectors orbitting a rotating BTZ black hole of angular momentum $J/M\ell=0.9999$. In each plot, the UDW detectors are have a proper separation of $d(R_A,R_B)=1$, the energy gap is $\Omega\sigma=1$, the black hole, has a mass of $M=10^{-3}$, $\zeta=1$ and $\ell/\sigma=10$.}
        \label{fig: PX0.001}
\end{figure*}

The origin of this entanglement amplification as the black hole approaches extremality is quite subtle. It is not due to any divergences in the respective coefficients  $K_P$ and  $K_X$ of detector probability and non-locality, both of which are independent of $r_-$ and thus are well-behaved as $r_+\to r_-$ (see Supplemental Material).
It is also not a kinematic effect of the type found in
AdS-Rindler spacetime \cite{jennings2010response,Henderson:2019uqo}, as an investigation of the partial image sum reveals. In Figure \ref{fig: partial sum}, we plot the detector probabilities, non-locality, and concurrence for several different partial image sums (whose general expressions we provide in the Supplementary Material). There is no  entanglement amplification for the $n=0$ term, which corresponds to the AdS-Rindler case \cite{Henderson:2019uqo}.  Instead, as the number of terms in the sums increases, the peak in $P_A$ shifts to the right and the peak in $X$ tends to broaden;  the peak in the concurrence becomes evident once $|n|>3$. Close to the horizon and at intermediate distances, the sum rapidly converges for $|n| > 30$; indeed,
the difference between $\sum_{n=-1}^1$ and $\sum_{n=-7}^7$ is greater than the difference between $\sum_{n=-7}^7$ and $\sum_{n=-100}^{100}$.
Plotting $P_A$, $P_B$, and $|X|$ in figure~\ref{fig: PX0.001}, we see that there is  a monotonic increase in $|X|$ with increasing $J$ 
in the intermediate zone about $20\sigma < d(r_+,R_A) < 40\sigma$, whilst both   $P_A$ and $P_B$ reach a maximum and then begin to decrease, as illustrated in  figure~\ref{fig: PX0.001}.
The spacetime thus has relatively enhanced non-local correlations in regions far, but not too far, for near-extremal black holes.

We also find that the  ergosphere plays no significant role in harvesting. Requiring $-(N^\perp)^2 + r^2 (N^\phi)^2 =  M - r^2/L^2 > 0$, the ergosphere extends out to $R=\sqrt{M}\ell$. The entanglement shadow extends beyond the ergosphere for the cases of $\Omega\sigma=0.01$ and $\Omega\sigma=0.1$, though lies entirely within the ergosphere for $\Omega\sigma=1$.  From  figure~\ref{fig: PX0.001} we see that as detector $A$ approaches the horizon its response $P_A$ increases to a maximum before sharply dropping off. This takes place for all values of $J$, and is signatory of the anti-Hawking effect recently observed for static black holes \cite{Henderson2019b}.  We find for larger gaps that $P_A$ experiences a sharp increase near the horizon before decreasing.

What of other vacuum correlations? An investigation of the mutual information 
\begin{equation}
  I_{AB} = \mathcal{L}_+ \log \mathcal{L}_+  + \mathcal{L}_- \log \mathcal{L}_-  - {P}_A  \log{P}_A  -{P}_B  \log{P}_B 
\end{equation}
where $ \mathcal{L}_\pm = \frac{1}{2}\left({P}_A   + {P}_B   \pm \sqrt{({P}_A  -{P}_B )^2+ 4 |L_{AB} |^2 } \right)$ 
indicates that it  peaks away from the black hole, 
albeit at smaller distances than the concurrence, and likewise undergoes significant amplification as extremality is approached.  The nearer location of the peak
occurs because  $P_A$, $P_B$, and $|L_{AB}|$ are all sharply peaked relatively close to the horizon, whereas $|X|$ peaks at notably larger distances. Hence  only the quantum correlations are enhanced in the intermediate regime.

{\it Conclusion}\;
\label{sec: conclusion}
We have shown that near-extremal BTZ black holes amplify the amount of quantum vacuum entanglement that can be extracted by a pair of idealized detectors.
Remarkably the concurrence can reach a maximum value as large as 10 times that of its static counterpart, depending on the energy gap of the detectors, 
and occurs at intermediate distances from the black hole. Significant harvesting is possible at large spacelike detector separations.  This is due to a diminishment of detector noise at such distances  relative to non-local correlations, which monotonically increase for increasing $J$ at any fixed distance within this intermediate region.

The effects we find are present for small mass black holes.   For larger mass black holes we find that the quantities $P_A$, $P_B$, and $|X|$ (and therefore $\mathcal{C}$) are increasingly insensitive to the angular momentum, and for $M\geq 1$, the sensitivity is vanishingly small.  This is because the additional terms in the image sum of equation (\ref{eq: sum}) become very tiny, and the physics is dominated by AdS-Rindler effects described previously \cite{Henderson2019b}. 

It is quite remarkable that the effects of rotation can amplify vacuum correlations so dramatically.  How this effect persists for other black holes remains an interesting subject for future study. \\

{\it Acknowledgements}
$\quad$ 
We thank Erickson Tjoa and Jorma Louko their useful comments and discussions. MR and LJH were funded by Natural Science and Engineering Research Council of Canada (NSERC) graduate scholarships. This research was supported in part by NSERC and the Perimeter Institute for Theoretical Physics. Research at Perimeter Institute is supported in part by the Government of Canada through the Department of Innovation, Science and Economic Development Canada and by the Province of Ontario through the Ministry of Colleges and Universities.

\bibliographystyle{unsrt}
\bibliography{EntanglementHarvestingRefs}

\onecolumngrid
\appendix

\section{Supplementary material: calculations for detector probabilities, non-localities, and correlations}

\subsection{Probabilities and Non-local Correlations}
In general, the detector probabilities ($P_D$) and non-locality term $X$ are given by
\begin{align}
P_D&=\lambda^2\int d\tau_Dd\tau_D'\chi_D(\tau_D)\chi_D(\tau_D')e^{-i\Omega_D(\tau_D-\tau_D')}W(x_D,x_D(\tau_D'))\label{eq: PD}\\
X&=-\lambda^2\int d\tau_Ad\tau_B\chi(\tau_A)\chi(\tau_B)e^{-i(\Omega_A\tau_A+\Omega_B\tau_B)}\left[\Theta[t-t']W(x_A(t),x_B(t'))+\Theta[t-t']W(x_B(t'),x_A(t))\right] \label{eq: X}
\end{align}
where $\chi$ is the switching function, $\Omega_D$ is the energy gap of detector $D$, $x_D$ is the trajectory of detector D, $\Theta$ is the Heaviside step function, $\lambda\ll1$ is the file/director coupling constant, and $W$ is the Wightman function,
\begin{align}
W_{BTZ}=\frac{1}{4\pi}\frac{1}{\sqrt{2}\ell}\lim_{\epsilon\to 0}\sum_{n=-\infty}^{n=\infty}\eta^n\left(\frac{1}{\sqrt{\sigma_{\epsilon}(x,\Gamma^nx')}}-\frac{\zeta}{\sqrt{\sigma_{\epsilon}(x,\Gamma^nx')+2}}\right)
\end{align}
where
\begin{equation}
\begin{aligned}
\sigma_\epsilon(x,\Gamma^nx')=&-1+\sqrt{\alpha(r)\alpha(r')}\cosh\left[\frac{r_+}{\ell}(\Delta\phi-2\pi n)-\frac{r_-}{\ell^2}(t-t'-i\epsilon)\right]\\
&-\sqrt{(\alpha(r)-1)(\alpha(r')-1)}\cosh\left[\frac{r_+}{\ell^2}(t-t'-i\epsilon)-\frac{r_-}{\ell}(\Delta\phi-2\pi n)\right]
\label{eq: sigma 0}
\end{aligned}
\end{equation}
is the squared geodesic distance between any two points (which can be determined by radar ranging in any frame),
and
\begin{align}
\alpha(r)&=\frac{r^2-r_-^2}{r_+^2-r_-^2}\qquad
\Delta\phi=\phi-\phi'
\end{align}
 We calculate these quantities by assuming Gaussian switching: $\chi(\tau_A)=e^{-\tau_A^2/2\sigma_D^2}$ and $\chi(\tau_B)=e^{-\tau_B^2/2\sigma_D^2}$. For simplicity, we also take the width $\sigma_D$ of the two switching functions to be the same as well as the energy gap $\Omega_A=\Omega_B=\Omega$. and we work in the co-rotating frame:\begin{align}
t&=\frac{\ell r_+\tau}{\sqrt{r^2-r_+^2}\sqrt{r_+^2-r_-^2}} \label{eq: CRM t}\\
\phi&=\frac{r_-\tau}{\sqrt{r^2-r_+^2}\sqrt{r_+^2-r_-^2}} \label{eq: CRM phi}\ .
\end{align}
Through straightforward though tedious manipulations, we find $P_D=\sum_{n=-\infty}^\infty\eta^n\left\{I_n^--\zeta I_n^+\right\}$, where
\begin{align}
I_n^\pm=K_P\int_{-\infty}^{\infty}dz\frac{e^{-a\left(z-\frac{2\pi nr_-}{\ell}\right)^2}e^{-i\beta\left(z-\frac{2\pi nr_-}{\ell}\right)}}{\sqrt{\left(\cosh(\alpha_n^\pm)-\cosh\left[z\right]\right)}}
\label{eq: In}
\end{align}
and
\begin{align}
K_P&=\frac{\lambda ^2 \sigma_D}{4 \sqrt{2 \pi }}\\
a&=\left(\frac{R^3}{r_+^2-r_-^2}\right)^2/4\sigma_D^2\\
\beta&=\frac{\Omega_DR^3}{r_+^2-r_-^2}\\
\cosh(\alpha_n^\pm)&=\frac{1}{\alpha(r)-1}\left[\pm1+\alpha(r)\cosh\left(2\pi n\frac{r_+}{\ell}\right)\right]
\end{align}

To calculate the non-locality $X$, let
\begin{align}
x_A(\tau_A)&:=\left\{t=\frac{\ell r_+\tau_A}{\sqrt{r^2-r_+^2}\sqrt{r_+^2-r_-^2}},r=R_A,\phi_A=\frac{r_-\tau_A}{\sqrt{r^2-r_+^2}\sqrt{r_+^2-r_-^2}}\right\}\\
x_B(\tau_B)&:=\left\{t'=\frac{\ell r_+\tau_B}{\sqrt{r^2-r_+^2}\sqrt{r_+^2-r_-^2}},r=R_B,\phi_B=\frac{r_-\tau_B}{\sqrt{r^2-r_+^2}\sqrt{r_+^2-r_-^2}}\right\}
\end{align}
Similarly tedious calculations show $X=\sum_{n=-\infty}^\infty\eta^n\left[\left(I^-_{AB,n}+I^-_{BA,-n}\right)-\zeta\left(I^+_{AB,n}+I^+_{BA,-n}\right)\right]$ where
\begin{align}
I_{AB,n}^\pm+I_{BA,-n}^\pm=\frac{K_X}{2}\int_0^\infty dz\left[\frac{e^{-a_X\left(z-\frac{2\pi nr_-}{\ell}\right)^2}e^{-i\beta_X\left(z-\frac{2\pi nr_-}{\ell}\right)}+e^{-a_X\left(z+\frac{2\pi nr_-}{\ell}\right)^2}e^{i\beta_X\left(z+\frac{2\pi nr_-}{\ell}\right)}}{\sqrt{\cosh(\alpha_n^\pm)-\cosh z}}\right]
\end{align}
with
\begin{align}
K_X&=-\frac{\lambda ^2 \sigma _A \sigma _B \sqrt[4]{\left(R_A^2-r_+^2\right)
   \left(R_B^2-r_+^2\right)} \exp \left(-\frac{\sigma _A^2 \sigma _B^2 \left(\Omega _A
   \sqrt{R_A^2-r_+^2}+\Omega _B \sqrt{R_B^2-r_+^2}\right){}^2}{2 \left(\sigma _A^2
   \left(R_B^2-r_+^2\right)+\sigma _B^2 \left(R_A^2-r_+^2\right)\right)}\right)}{2
   \sqrt{\pi } \sqrt{\sigma _A^2 \left(R_B^2-r_+^2\right)+\sigma _B^2
   \left(R_A^2-r_+^2\right)}}\\
a_X&=\frac{1}{2(\bar\sigma_A^2+\bar\sigma_B^2)}\left(\frac{\ell^2 r_+}{r_+^2-r_-^2}\right)^2\\
\beta_X&=\frac{(\bar\sigma_B^2\bar\Omega_B-\bar\sigma_A^2\bar\Omega_A)}{\bar\sigma_A^2+\bar\sigma_B^2}\left(\frac{\ell^2 r_+}{r_+^2-r_-^2}\right)\\
\end{align}
and
\begin{equation}
\begin{aligned}
\bar\sigma_A&=\sigma_A/\gamma_A\\
\bar\sigma_B&=\sigma_B/\gamma_B\\
\Omega_A\tau_A&=\Omega_A\gamma_At:=\bar\Omega_A t\\
\Omega_B\tau_B&=\Omega_B\gamma_Bt:=\bar\Omega_B t
\end{aligned}
\end{equation}

\subsection{Detector Correlations}

The detector correlations are given by 

\begin{align}
L_{AB}=\lambda^2\int dtdt'\eta_B(t')\eta_A(t)e^{-i[\Omega_B\tau_B(t')-\Omega_A\tau_A(t)]}W(x_A(t),x_B(t'))\ ,
\label{eq: C}
\end{align}
where $\eta_D(t)=\frac{d\tau_D}{dt}\chi(\tau_D)=\gamma_De^{-\tau_D^2/2\sigma_D^2}$ and
\begin{equation}
\begin{aligned}
\gamma_A&=\frac{\sqrt{R_A^2-r_+^2}\sqrt{r_+^2-r_-^2}}{\ell r_+}\\
\gamma_B&=\frac{\sqrt{R_B^2-r_+^2}\sqrt{r_+^2-r_-^2}}{\ell r_+}
\end{aligned}
\end{equation}
Working in the co-rotating frame, we can write $L_{AB}=\sum_{n=-\infty}^\infty\left(I_n^--\zeta I_n^+\right)$ where
\begin{align}
I_n^\pm=K_L\int_{-\infty}^\infty dz \frac{e^{-a_L(z-2\pi nr_-/\ell)^2}e^{-i\beta_L(z-2\pi nr_-/\ell)}}{\sqrt{\cosh(\alpha_n^\pm)-\cosh\left[z\right]}}
\end{align}
with
\begin{align}
K_L&=\frac{1}{\sqrt{\sqrt{\alpha(R_A)-1}\sqrt{\alpha(R_B)-1}}}\frac{\lambda^2}{8\pi\sqrt{2}\ell}\gamma_B\gamma_A\left(\frac{2\sqrt{2\pi}\bar\sigma_A\bar\sigma_B}{\sqrt{\bar\sigma_A^2+\bar\sigma_B^2}}\right)e^{-\frac{\bar\sigma_A^2\bar\sigma_B^2(\bar\Omega_A-\bar\Omega_B)^2}{2(\bar\sigma_A^2+\bar\sigma_B^2)}}\frac{\ell^2r_+}{r_+^2-r_-^2}\\
a_L&=\frac{1}{2(\bar\sigma_A^2+\bar\sigma_B^2)}\left(\frac{\ell^2 r_+}{r_+^2-r_-^2}\right)^2\\
\beta_L&=\frac{(\bar\sigma_B^2\bar\Omega_B+\bar\sigma_A^2\bar\Omega_A)}{\bar\sigma_A^2+\bar\sigma_B^2}\left(\frac{\ell^2 r_+}{r_+^2-r_-^2}\right)
\end{align}

\begin{figure}[h]
\centering
\begin{tikzpicture}
  \draw[thick, ->] (-6,0) -- (6,0) coordinate (xaxis);

  \draw[thick, ->] (0,0) -- (0,4) coordinate (yaxis);

  \node[above] at (xaxis) {$\mathrm{Re}(\tau_B)$};

  \node[right]  at (yaxis) {$\mathrm{Im}(\tau_B)$};

\begin{scope}[very thick,decoration={
    markings,
    mark=at position 0.5 with {\arrow[line width=3pt]{<}}}
    ] 
    \draw[postaction={decorate}] (4,0)--(4,0) node[below, black] {$5s$};
    \draw[blue, postaction={decorate}] (4,0)--(4,2) node[above, black] {$5s+i$};
    \draw[blue, postaction={decorate}] (4,2)--(-4,2) node[above, black] {$-5s+i$};
    \draw[blue, postaction={decorate}] (-4,2)--(-4,0) node[below, black] {$-5s$};
\end{scope}
\end{tikzpicture}
\caption{Contour to integrate over the $\tau_B$ variable when calculating $L_{AB}$.}
\label{fig: contour}
\end{figure}
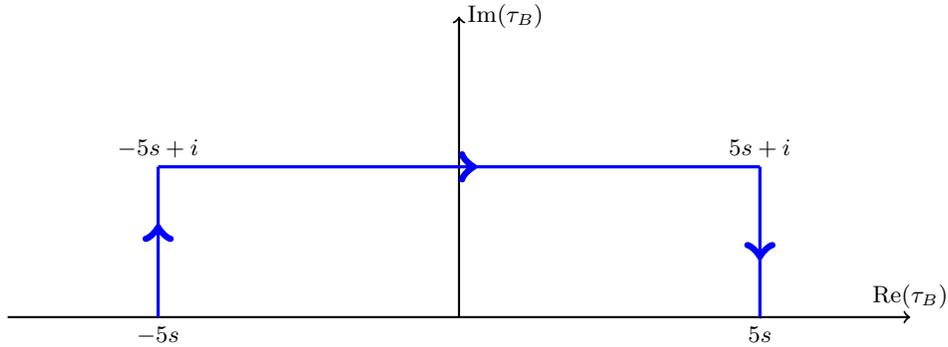

To calculate   detector correlations, equation \eqref{eq: C} was applied by using the co-rotating frame in conjunction with the methodology discussed in Ref [30] to a precision of $10^{-4}$. We approximated the integral as $-5s<\tau_A,\tau_B<5s$ and integrated $\tau_B$ on the contour shown in Figure \ref{fig: contour}.

\end{document}